\documentclass{pasj00}
\begin{document}
\SetRunningHead{Meng, Li and Yang}{The effect of metallicity on
the DTD of SNe Ia}
\Received{2000/12/31}%{yyyy/mm/dd}
\Accepted{2001/01/01}%{yyyy/mm/dd}

\title{The effect of metallicity on the delay-time distribution of type
Ia supernova}

%%% begin:list of authors
% Do NOT capitalize all letters in "textsc".
\author{X.-C. \textsc{Meng}$^{\rm 1}$ %
  \thanks{We are grateful to Dr. Richard Pokorny for improving the English
language of the original manuscript and thanks the anonymous
referee for his/her constructive suggestions which make the
manuscript more complete. This work was partly supported by
Natural Science Foundation of China under grant nos. 10963001 and
11003003, the Project of Science and Technology from the Ministry
of Education (211102), and the Project of the Fundamental and
Frontier Research of Henan Province under grant no. 102300410223.}
}
%\affil{Department of Physics
%and Chemistry, Henan Polytechnic University, Jiaozuo, 454000,
%China}
%\email{conson859@msn.com}

%\author{W.-M. \textsc{Yang}}
\affil{$^{\rm 1}$School of Physics and Chemistry, Henan
Polytechnic University, Jiaozuo, 454000,
China}\email{xiangcunmeng@hotmail.com} \and
\author{Z.-M. {\sc Li}$^{\rm 2, 3}$}
\affil{$^2$ Institute for Astronomy and History of Science and
Technology, Dali University, Dali, 671003, China,\\
$^3$ National Astronomical Observatories, Chinese Academy of
Sciences, Beijing, 100012, China}%\email{ccccc@xxx.xxx.xx.xx}
 \and
\author{W.-M. {\sc Yang}$^{\rm 1,4}$}
\affil{$^4$ Department of Astronomy, Beijing Normal University,
Beijing 100875, China}
%%% end:list of authors

%%% Please use the following style in case that sorting by
%%% affilation is impossible.
%
% \author{%
%   D-Firstname \textsc{D-Familyname}\altaffilmark{1}
%   E-Firstname \textsc{E-Familyname}\altaffilmark{1,2}
%   and
%   F-Firstname \textsc{F-Familyname}\altaffilmark{2}}
% \altaffiltext{1}{Address of Institute}
% \email{ddddd@xxx.xxx.xx.xx}
% \email{eeeee@xxx.xxx.xx.xx}
% \altaffiltext{2}{Address of Institute}

%% `\KeyWords{}' always has to be placed before `\maketitle'.
\KeyWords{stars: binaries: general---stars: supernovae: general---stars: white dwarfs} %Do NOT move this preamble from here!

\maketitle

\begin{abstract}
Measuring the delay-time distribution (DTD) of type Ia
supernova(SNe Ia) is an important way to constrain the progenitor
nature of SNe Ia. Recently, Strolger et al. (2010) obtained a very
delayed DTD, which is much different from other measurements. They
suggested that metallicity could be the origin of their delayed
DTD. In this paper, we show the effect of metallicity on the DTD
of SNe Ia from single-degenerate models (including WD + MS and WD
+RG channels). Via a binary population synthesis approach, we find
that the DTD from a low metallicity population is significantly
delayed compared with that from a high metallicity one. In
addition, we also find that a substantial fraction of SNe Ia have
a delay time shorter than 1 Gyr, and the fraction of SNe Ia with
short delay times increases with metallicity, i.e. about 35\% for
Z=0.001, while more than 70\% for Z=0.02. These results would help
to qualitatively explain the result of Strolger et al. (2010).
Furthermore, we noticed that the contribution of WD + RG channel
from the low metallicity population is higher than that from the
high metallicity one. However, we can not quantitatively obtain a
DTD consistent with the results of Strolger et al. (2010) by
changing metallicity.  As a consequence, metallicity may partly
contribute to the DTD of SNe Ia and should therefore be checked
carefully when one derives the DTD of SNe Ia from observations.
\end{abstract}

\section{Introduction}
\label{sect:1}
 Although Type Ia supernovae (SNe Ia) are very important in
cosmology (\cite{RIE98}; \cite{PER99}), the exact nature of their
progenitors is still unclear (\cite{HN00}; \cite{LEI00};
\cite{PAR07}). There is a consensus that SNe Ia result from the
thermonuclear explosion of a carbon--oxygen white dwarf (CO WD) in
a binary system (\cite{HF60}). According to the nature of the
companions of the mass accreting WDs, two basic scenarios for the
progenitors of SN Ia have been discussed over the last three
decades. One is the single degenerate (SD) model (\cite{WI73};
\cite{NTY84}), i.e. the companion is a main-sequence or a slightly
evolved star (WD+MS), or a red giant star (WD+RG) or a helium star
(WD + He star) (\cite{LI97}; \cite{HAC99a}; \cite{LAN00};
\cite{HAN04}; \cite{CHENWC07}; \cite{MENG09}; \cite{LGL09};
\cite{WANGB09a}; \cite{WANGB10}). The other is the double
degenerate (DD) model, i.e. the companion is another WD
(\cite{IT84}; \cite{WEB84}). Measuring the delay-time distribution
(DTD, delay time is the elapsed time between primordial system
formation and explosion as a SN Ia event) is a very important way
to distinguish between the different progenitor systems. Recently,
using the high-redshift SNe Ia sample ($0.2<z<1.8$) from the
Hubble Space Telescope ACS imaging of the GOODS North and South
fields, \citet{STROLGER10} showed a significantly delayed DTD that
is confined to 3-4 Gyr, which is difficult to resolve with any
intrinsic DTD. This result confirmed their previous findings
(\cite{STROLGER04}). But, they also noticed that this result is
mainly motivated by the decline in the number of SNe Ia at
$z>1.2$. Their sub-samples with low redshift ($z<1.2$) showed
plausible DTDs dominated by SNe Ia with short delay times. The
difference between their low-z and high-z results may be partly
explained by the fact that a substantial fraction of $z>1.2$
supernova may be obscured by dust. However, the DTD derived by
\citet{STROLGER10} may be dominated by systematic errors, in
particular due to uncertainties in the star formation history
(SFH, \cite{FORSTER06}). The inferred delay time is strongly
dependent on the peak in the assumed SFH and none of the popular
progenitor models under consideration can be ruled out with any
significant degree of confidence (see also \cite{ODA08} and
\cite{VALIANTE09}).

Moreover, the results of \citet{STROLGER10} are inconsistent with
many low and moderate redshift measurements, which showed that
most SNe Ia have delay times between 0.3 and 2 Gyr
(\cite{SCHAWINSKI09}), and there are also SNe Ia with very long
delay times (older than 10 Gyr inferred from SNe Ia in elliptical
galaxies in the local universe, \cite{MAN05}) or extremely short
delay times (shorter than 0.1-0.3 Gyr, \cite{MAN06};
\cite{SCHAWINSKI09}; \cite{RASKIN09}). Based on some observations,
i.e. the strong enhancement of the SN Ia birthrate in radio-loud
early-type galaxies, the strong dependence of the SN Ia birthrate
on the colors of the host galaxies, and the evolution of the SN Ia
birthrate with redshift (\cite{DEL05}; \cite{MAN05};
\cite{STROLGER04}), \citet{MAN06} suggested a bimodal DTD, in
which some of the SNe Ia explode soon after starburst with a delay
time less than 0.1-0.5 Gyr (`prompt' SNe Ia, \cite{SCHAWINSKI09};
\cite{RASKIN09}), while the rest have a much wider distribution
with a delay time of about 3 Gyr (`tardy' SNe Ia). In theory, the
bimodal DTD may be constructed from detailed binary population
synthesis (\cite{MENGYANG10a}). However, the excess of SNe Ia in
radio galaxies is the only one observation that strongly indicates
an extremely large amount of the prompt population and hence is
distinct from the longer delay time population (see \cite{MAN06}),
\textbf{but} this excess is not supported by a more recent
observation (\cite{GRAHAM10}). By comparing with host galaxy
color, some authors proposed a simple two-component model, A+B
model, which may be a variation of the bimodal DTD. (\cite{SB05};
\cite{SULLIVAN06}; \cite{BRANDT10}). Recently, more and more
observational evidence showed that the DTD of SNe Ia follows the
power-law form of $t^{\rm -1}$ which is much different from the
results of \citet{STROLGER10}. The power-law form is even
different from the bimodal model or the A+B model, which might
indicate that the simple two-component model is an insufficient
description for observational data. (\cite{TOTANI08};
\cite{MAOZ10b}; \cite{MAOZ10a, MAOZ10}). Even via the same method
as \citet{STROLGER10}, i.e. comparison between cosmic SFR
evolution and SN Ia rate evolution, a $t^{\rm -1}$ DTD was also
found to be in nice agreement with observed data (\cite{GRAUR11}).
The DTD derived by \citet{STROLGER10} is well confined to 3-4 Gyr
which is strongly inconsistent with those DTDs mentioned above,
and only a small fraction belong to the ``prompt''\footnote{The
delay time of the prompt SNe Ia in \citet{STROLGER10} is still
much delayed compared with that suggested by \citet{MAN06}} SNe Ia
population, i.e. smaller than 10\%. However, some low-redshift
samples show the existence of prompt SNe Ia at a high confidence
level, and the birth rate of the prompt component is much higher
than that of the tardy SNe Ia (\cite{AUBOURG08}; \cite{MAOZ10}).
Theoretically, short delay SNe Ia may also be produced by a WD +
helium star and WD + MS channel (\cite{WANGB09a};
\cite{MENGYANG10a}).

Whatever, \citet{STROLGER10} suggested that the effect of
metallicity could be one possible resolution for disagreement
between their discovery and low/moderate results. In this paper,
we want to check whether changing the metallicity can create DTDs
consistent with the results of \citet{STROLGER10}, in the
framework of the single-degenerate scenario.

In section \ref{sect:2}, we simply describe our method, and
present the calculation results in section \ref{sect:3}. In
section \ref{sect:4}, we show discussions and our main
conclusions.

% Authors can give a citation as `Michel et al. 1992'.
% You may also use \cite, \citep and \citet for citation, and use Table~1
% or Figure~1 and so forth. Using \ref and \label for cross-references of
% Tables/Figures is a good way in adjusting/adding/removing text, tables or
% figures.

%\noindent IMPORTANT NOTICE\\
%1. ``\verb|\draft|'' creates single column and double spaces format.\\
%2. If you comment out ``\verb|\draft|'', the output will be double column
%   and single space.\\
%3. For cross-references, the use of ``\verb|\label|, \verb|\ref|, \verb|\cite|''
%   and the thebibliography environment is strongly recommended. \\
%4. Do NOT use ``\verb|\def|, \verb|\renewcommand|''.\\
%5. Do NOT redifine commands provided by PASJ00.cls.\\

%\newpage

\section{METHOD}\label{sect:2}
Recently, \citet{MENGYANG10a} constructed a comprehensive single
degenerate progenitor model for SNe Ia. In this model, the
mass-stripping effect of optically thick wind (\cite{HAC96}) and
the effect of a thermally instable disk were included
(\cite{HKN08}; \cite{XL09}). The prescription of \citet{HAC99a} on
WDs accreting hydrogen-rich material from their companions was
applied to calculate the WD mass growth. The optically thick wind
and the material stripped-off by the wind were assumed to take
away the specific angular momentum of the WD and its companion,
respectively. In \citet{MENGYANG10a}, both WD + MS channel and WD
+ RG channel are considered, i.e. Roche lobe overflow (RLOF)
begins at MS or RG stage. The Galactic birth rate of SNe Ia
derived from that model is comparable with that from observations.
In addition, this model may even explain some supernovae with low
hydrogen mass in their explosion ejecta (\cite{MENGYANG10b}).
\citet{MENGYANG10a} calculated more than 1600 WD close binary
evolutions and showed the initial parameter space leading to SNe
Ia in an orbital period - secondary mass ($\log P_{\rm i}, M_{\rm
2}^{\rm i}$) plane, and these results may be conveniently used in
a binary population synthesis code for obtaining the DTD of SNe
Ia.

\begin{figure}
  \begin{center}
%    \FigureFile(80mm,80mm){mms02120.ps}
    \includegraphics[width=6.cm, angle=270]{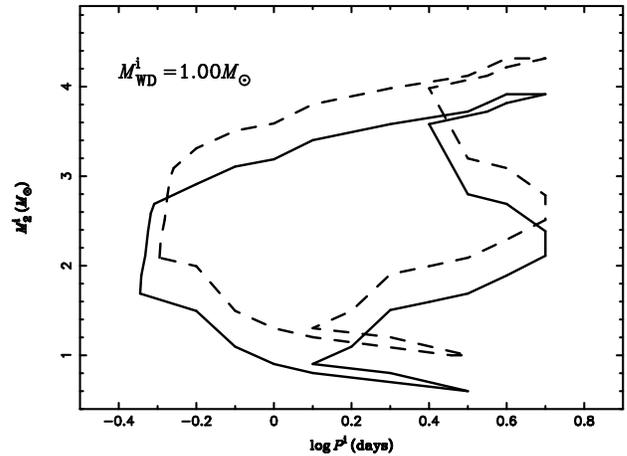}
    %%% \FigureFile(width,height){filename}
  \end{center}
  \caption{The contours shown by solid and dashed lines are the parameter space for SNe Ia with $Z=0.001$
  and $Z=0.02$, respectively, where the initial WD mass is $1.00 M_{\odot}$.} \label{move}
\end{figure}

The delay time of a SN Ia from the SD model is mainly determined
by the stellar evolutionary timescale of the secondary, and thus
the secondary mass. That is to say that the DTD of SNe Ia is a
function of the location of the parameter space in the ($\log
P_{\rm i}, M_{\rm 2}^{\rm i}$) plane. However, this location is
directly affected by metallicity. For a system with given initial
WD mass and orbital period, the initial mass of the companion
leading to SNe Ia increases with metallicity, i.e. the upper
boundary and lower boundary of the companion mass move to lower
values with the decrease of metallicity (see the figure 4 in
\cite{MENG09}). Thus, the DTD of SNe Ia are affected by
metallicity via companion mass. Between the two boundaries of the
companion mass for SNe Ia, the lower boundary dominates the longer
delay time of SNe Ia. The low boundary is mainly determined by the
condition that the mass transfer rate between a CO WD and its
companion is higher than a critical value which is the lowest
accretion rate of a CO WD avoiding violent nova explosion, while
the upper boundaries are mainly determined by dynamically instable
mass transfer and the strong hydrogen-shell flash. The mass
transfer rate for a given binary system is closely related to
metallicity, which is due to the correlation between stellar
structure and metallicity (\cite{UME99}; \cite{CHE07}). Generally,
the time-scale for mass transfer is the thermal time-scale, which
increases with metallicity. This leads to a higher mass-transfer
rate for a low metallicity system (\cite{LAN00}). So, low-mass
companions with low metallicity are thus more likely to fulfills
the constraint for mass transfer than those with high metallicity.
A WD + MS system with a low metallicity is therefor more likely to
be the progenitor of a SN Ia (see also \cite{MENG09}). On the
other hand, a high mass-transfer rate means that a binary system
with the same initial parameters but al low metallicity will
fulfill the condition of dynamical instability more possible. Even
though the mass transfer for the system is dynamically stable, the
mass-transfer rate could be so high that most of the transferred
material is lost from the system by the optically thick wind, and
at the same time, a large amount of hydrogen-rich material
stripped off by the wind is lost from the companion envelope. The
mass-transfer rate then sharply decreases to less than the
critical value for avoiding the strong hydrogen-shell flash after
mass-ratio inversion. As a consequence, the initial parameter
space for SNe Ia moves to a lower companion mass with the decrease
of metallicity (\cite{MENG09}). For example, for a system with
given initial WD mass and initial orbital period, the companion
mass for $Z=0.001$ is lower than that of $Z=0.02$ by about 0.4
$M_{\odot}$ (see also \cite{CHENWC09}). In this paper, we do not
calculate binary evolution for low metallicity, instead we move
the parameter space for SNe Ia of $Z=0.02$ given by
\citet{MENGYANG10a} to a lower companion mass by 0.4 $M_{\odot}$
and assume rather arbitrarily that the parameter space with a low
companion mass is equivalent to that for $Z=0.001$. To clarify
what we did, we use the case of $M_{\rm WD}^{\rm i}=1.00M_{\odot}$
as an example (see figure \ref{move})\footnote{In this paper, we
do not directly use the results in \citet{MENG09} because the
calculations in the paper is not as complete as in
\citet{MENGYANG10a}, i.e. they did not consider the WD + RG
channel which is the dominant one for old SNe Ia, and did not
incorporate some special effects such as the mass-stripping effect
of optically thick wind and the effect of a thermally instable
disk which could be very important for SNe Ia.}. Since we only
want to check whether the metallicity has an ability to create a
DTD matching with the discovery of \citet{STROLGER10}, this simple
assumption is not unreasonable (see discussion in section
\ref{sect:4.3}).

To obtain the DTD of $Z=0.001$, we have performed a series
detailed Monte Carlo simulations via Hurley's rapid binary
evolution code (\cite{HUR00, HUR02}). In the simulations, if a
binary system evolves to a WD + MS or WD + RG stage, and the
system is located in the ($\log P^{\rm i}, M_{\rm 2}^{\rm i}$)
plane for SNe Ia at the onset of RLOF, we assume that a SN Ia is
produced. In the simulations, we follow the evolution of $10^{\rm
7}$ sample binaries. The evolutional channel is described in
\citet{MENGYANG10a}. As in \citet{MENGYANG10a}, we adopted the
following input for the simulations. (1) A single starburst is
assumed, i.e. $10^{\rm 11} M_{\odot}$ in stars are produced at one
time. (2) The initial mass function (IMF) of \citet{MS79} is
adopted. (3) The mass-ratio distribution is taken to be constant.
(4) The distribution of separations is taken to be constant in
$\log a$ for wide binaries, where $a$ is the orbital separation.
(5) A circular orbit is assumed for all binaries. (6)The common
envelope (CE) ejection efficiency $\alpha_{\rm CE}$, which denotes
the fraction of the released orbital energy used to eject the CE,
is set to 1.0 or 3.0. (See \citet{MENGYANG10a} for details).

%n figure \ref{fig:sample}, ...

%\begin{figure}
%  \begin{center}
%    \FigureFile(80mm,80mm){fig1.eps}
%    %%% \FigureFile(width,height){filename}
%  \end{center}
%  \caption{This is the first figure.}\label{fig:sample}
%\end{figure}

\section{RESULT}\label{sect:3}

\begin{figure}
  \begin{center}
%    \FigureFile(80mm,80mm){mms02120.ps}
    \includegraphics[width=6.cm, angle=270]{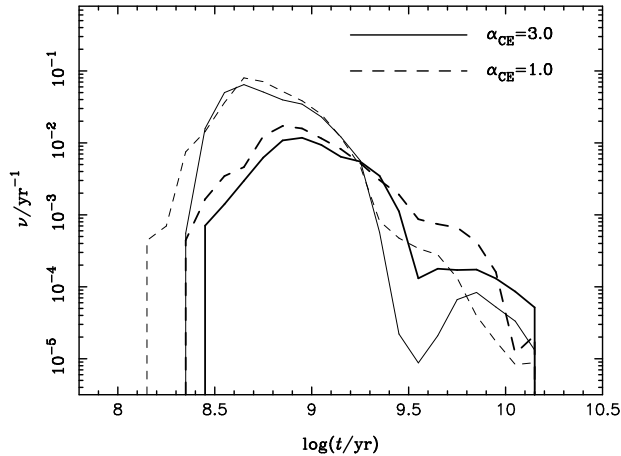}
    %%% \FigureFile(width,height){filename}
  \end{center}
  \caption{Evolution of the birthrates of SNe Ia for a single
starburst of $10^{\rm 11}M_{\odot}$ for different $\alpha_{\rm
CE}$ (solid lines: $\alpha_{\rm CE}=3.0$; dashed lines:
$\alpha_{\rm CE}=1.0$) and different metallicities. The thick
lines are the results for $Z=0.001$ in this paper and the thin
ones are for $Z=0.02$ from \citet{MENGYANG10a}.} \label{001}
\end{figure}

In figure \ref{001}, we show the evolution of the birthrates of
SNe Ia for a single starburst for different $\alpha_{\rm CE}$ and
different metallicities. We can see from the figure that whatever
the $\alpha_{\rm CE}$, the DTDs of $Z=0.001$ are significantly
delayed compared with those of $Z=0.02$. For $Z=0.02$, SNe Ia
mainly occur between 0.2 and 2 Gyr with a mean value of 0.89 Gyr
after the burst, while they occur between 0.3 and 3.5 Gyr with a
mean value of 1.94 Gyr for $Z=0.001$. This is mainly derived from
the low companion mass for low metallicity. As stated in
\citet{HAN04} and \citet{MENGYANG10a}, we found that a high
$\alpha_{\rm CE}$ leads to a systematically later explosion time
for $Z=0.001$, because a high $\alpha_{\rm CE}$ leads to wider WD
binaries, and as a consequence, it takes a longer time for the
secondary to evolve to fill its Roche lobe. As noticed by
\citet{MENG09}, the peak value of the DTD for low metallicity is
lower than that for high metalicity and the WD + MS channel is the
dominant channel for the peak value. However, the contribution of
WD + RG channel to SNe Ia for $Z=0.001$ is higher than that for
$Z=0.02$, i.e. 1\%-2\% for $Z=0.02$, but 8.6\%-16\% for $Z=0.001$.
Actually, \citet{MENG09} noticed that the WD + RG channel may be
more common for low metallicity (see footnote 1 in \cite{MENG09}).

We also checked the fraction of SNe Ia with short delay times and
found that a substantial fraction of SNe Ia have a delay time
shorter than 1 Gyr and the fraction of SNe Ia with short delay
times increases with metallicity, i.e. about 35\% for $Z=0.001$,
while more than 70\% for $Z=0.02$.

\section{DISCUSSIONS AND CONCLUSIONS} \label{sect:4}
\subsection{comparison wiht observations}\label{sect:4.1}
Measuring the DTD of SNe Ia is an important way to constrain the
nature of the progenitor of SNe Ia. Recently, \citet{STROLGER10}
used data from the Hubble space telescope to confirm their
previous results that the data are largely inconsistent with
progenitor scenarios with short delay time, which is difficult to
explain with any intrinsic DTD. \citet{STROLGER10} suggested
\textbf{a} possible resolution for their results, i.e environment
such as metallicity may affect the progenitor mechanism
efficiently, especially in the early universe. Our results in this
paper seem to support this suggestion because we found that low
metallicity may significantly delay the DTD of SNe Ia. If the
result obtained by \citet{STROLGER10} shows the real nature of the
DTD of SNe Ia, metallicity could be an indispensable factor which
must be considered when studying the progenitors of SNe Ia.
However, for a high-z SNe Ia sample as used by \citet{STROLGER10},
whether metallicity works as significantly as suggested in this
paper should be checked carefully since optically thick wind could
not work, and thus SNe Ia could not occur in a low-metallicity
environment (\cite{KOB98}). Furthermore, the evolution of
metallicity with redshift should be considered carefully (see next
section), since the results of \citet{STROLGER10} depend strongly
on the SNe Ia sample with $z>1.2$. The DTD derived from the
sub-sample with $z<1$ is consistent with results from low-redshift
supernova samples, which are better reproduced by our model. Thus,
the properties of SNe Ia with $z>1.2$ (their progenitors may form
at a red shift as large as 3-4) are interesting.

Recently, By comparing a theoretical DTD and an observational one
from \citet{TOTANI08}, \citet{Mennekens10} claimed that the DTD
from the single degenerate model is incompatible with
observations, which is mainly derived from a lower birth rate of
SNe Ia at long delay time, especially at a time longer than 8 Gyr.
Metallicity may improve the situation of the SD model since a low
metallicity may increase the birth rate of SNe Ia from WD + RG
channel as suggested in this paper and after all, the DTD derived
from the SD model by \citet{MENGYANG10a} is not incompatible with
observations.

\subsection{the evolution of DTD with redshift}\label{sect:4.2}
Generally, the mean value of metallicity decreases slowly with
redshift. Based on the results in this paper, the mean delay time
of SNe Ia could thus increases with redshift. However, the
evolution of metallicity is not a simple, monotonic function of
redshift. Metallicity shows a significant scatter at all redshifts
and the scatter increases with redshift (\cite{NAAGAMINE01}). At
$z\simeq3-4$, the mean value of metallicity is close to $0.005$ .
According to the results here, the progenitors of SNe Ia formed at
this redshift interval would then explode at an interval of
$z\simeq1.5-2.5$, and the delay time could be shorter than 2.75
Gyr. Even at $z=5$, the mean metallicity is $0.004$. The
progenitor stars formed at such high redshifts contribute to SNe
Ia exploding at $z>2$, which is beyond the the scope of SNe Ia
sample used by \citet{STROLGER10}. Stars formed at $z\simeq2$
usually have a metallicity lower than $0.02$, i.e. the metallicity
spreads from $0.01-1Z_{\odot}$ and has a mean value of
$\sim0.6Z_{\odot}$, where $Z_{\odot}$ is solar metallicity
(\cite{NAAGAMINE01}). Thus, the progenitors formed at $z\simeq2$
could not contribute to SNe Ia with long delay times at
$z\simeq1-2$, and the delay time at this redshift interval should
be shorter than 2 Gyr. For stars formed at $z\leq1.5$, the
distribution of metallicities is rather uniform, i.e. close to
$Z_{\odot}$, and these stars mainly contribute to SNe Ia exploding
at $z\leq1$, which means that SNe Ia at $z\leq1$ usually have a
delay time shorter than 2 Gyr. So, there is no time at which the
average metallicity of $Z<0.001$ contributs to the high-redshift
sample of \citet{STROLGER10}. Even if it were, our model would
predict $\sim35$\% of the SNe Ia to be ``prompt'', which is
inconsistent with the results of \citet{STROLGER10}. However, as
noticed by \citet{NAAGAMINE01}, the spread of metallicity
gradually increases toward high redshift. It takes $0.01-1.0
Z_{\odot}$ at $z=2$, but $10^{\rm -6}-3.0 Z_{\odot}$\footnote{The
existence of an optically thick wind is in doubt for very low
metallicity (\cite{KOB98}).} at $z>3$. Thus, there are still a few
progenitor stars formed at $z>3$ which could contribute to SNe Ia
with long delay times at $z\simeq1-2$. So, although metallicity
has no ability to create a DTD matching the discovery of
\citet{STROLGER10}, it still partly contributes to the discovery.

The WD + MS channel is the dominant one for SNe Ia, and the
discussion above is mainly based on this channel. Although the
contribution is small (1\% - 16\%, see section \ref{sect:3}), the
WD + RG channel contributes to SNe Ia with very long delay times,
i.e. longer than 3.5 Gyr. So the SNe Ia from this channel could
only be discovered at $z<2$, and the progenitor stars formed at
$z<4$ could contribute to SNe Ia at $z<1$.

\subsection{uncertainties}\label{sect:4.3}
Obviously, there exist many uncertainties for our discussions in
this paper. Firstly, we did not calculate the binary evolution
with low metallicity, i.e. we did not obtain the appropriate
parameter space for SNe Ia with low metallicity. Fortunately, some
works have referred to this problem, i.e. the parameter space for
SNe Ia moves to lower secondary mass with the decrease of
metallicity, and then a low metallicity leads to a longer delay
time (\cite{MENG09}; \cite{CHENWC09}). So, as the effect of
metallicity on the delay time of SNe Ia, we seem not to obtain a
new result. But, since our purpose is to check whether changing
metallicity can create DTDs consistent with the results of
\citet{STROLGER10}, our work is still meaningful. Then, the
following question is whether 0.4 $M_{\odot}$ used here is
reasonable although this value is obtained from detailed binary
evolution calculation. Actually, the lower boundary of the
parameter space for SNe Ia with $Z=0.001$ is lower than 0.8
$M_{\odot}$ when we move the parameter space for SNe Ia of
$Z=0.02$ to a lower companion mass by 0.4 $M_{\odot}$, which means
that we obtained an upper limit of the effect of metallicity on
the delay time of SNe Ia. In addition, a low metallicity reduces
the area of parameter space for SNe Ia (\cite{NOM99};
\cite{MENG09}; \cite{CHENWC09}). However, the area mainly affects
the birth rate of SNe Ia, not delay time, i.e. the birth rate of
$Z=0.001$ obtained here could be taken as an upper limit, but the
delay time here is still valid. So, the conclusion that
metallicity has no ability to interpret the observation of
\citet{STROLGER10} is reasonable.

Secondly, in the calculation of \citet{MENGYANG10a}, the optically
thick wind is assumed (\cite{HAC96}), which critically depends on
the opacity applied, i.e. it is likely that the wind does not work
when $Z<0.002$ and then SNe Ia should not be observed in
metal-poor environments (\cite{KOB98}). However, this prediction
was not uphold by some observations (\cite{PRIETO08};
\cite{BADENES09}; \cite{KHAN10}). To try to avoid arguments, we
assume that the wind still works when $Z=0.001$ (slightly lower
than the metalicity limit of $Z=0.002$, see also \cite{NOM99}),
and then the delay time obtained here should be taken as an upper
limit. Since the upper limit of delay time can not match with the
DTD of \citet{STROLGER10}, our conclusion is still hold no matter
which value of metallicity we choose.

Thirdly, the single-degenerate model in \citet{MENGYANG10a}
contains numerous assumptions, which may not be universally
accepted. For example, the mass-stripping effect of the optically
thick wind and the effect of the thermally instable disk were
included (\cite{HKN08}; \cite{XL09}). The mass stripping effect
mainly affects the birth rate of SNe Ia, i.e. the birth rate may
reduce significantly if the effect is not include (see
\cite{WANGB10}). The effect of the thermally instable disk affects
not only the birth rate, but also the delay time. If this effect
were not include, the birth rate and the delay time would decrease
significantly, which might lead to a DTD that is also not
consistent with the observation of \citet{STROLGER10}.

Finally, please keep in mind that there is an implicit assumption
in this paper that the assumptions used in \citet{MENGYANG10a} is
not affected by metallicity. This assumption is rather arbitrary
since great efforts are necessary to support it. Fortunately, some
previous studies showed that the influence of metallicity on the
assumptions used in SD model could be neglected. For example, the
critical accretion rate and the structure of WDs are almost not
affected by metallicity (\cite{MEN06}; \cite{UME99, UME99b}). So,
our assumption here
might not be a serious problem.\\

In summary, this paper fails its stated goal: to create a DTD
consistent with the measurement of \citet{STROLGER10} by changing
metallicity based on a SD scenario. Metallicity may only partly
resolve the long delay-time results of \citet{STROLGER10}.
However, when using the delay time derived from observations to
constrain the progenitors of SNe Ia, metallicity should be
carefully checked.

%%%
% See the manual for the detail.
%%%


\begin{thebibliography}{}
% Journals(e.g. A\&A,ApJ,AJ,NMRAS,PASP ...)
% Authors, Year, Journal, Vol#, Page#
% Journal Title Abbreviation >> http://www.asj.or.jp/pasj/Jabb.html

%\bibitem[Aldering et al.(2006)]{ALD06}
%Aldering G., Antilogus P., Bailey S. et al., 2006, ApJ, 650, 510
%\bibitem[Alexander \& Ferguson(1994)]{AF94}
%Alexander D. R., Ferguson J. W., 1994, ApJ, 437, 879
%\bibitem[\protect\citeauthoryear{Altavilla et al.}{2004}]{ALT04}
%Altavilla G. et al. 2004, MNRAS, 349, 1344
%\bibitem[Arnett(1982)]{ARN82}
%Arnett W.D., 1982, ApJ, 253, 785
%\bibitem[Arnett, Branch \& Wheeler(1985)]{ARN85}
%Arnett, W.D., Branch, D., Wheeler, J.C., 1985, Nature, 314, 337
\bibitem[Aubourg et al.(2008)]{AUBOURG08}
Aubourg \'{E}., Tojeiro R., Jimenez R. et al., 2008, A\&A, 492,
631
%\bibitem[(Baes et al.(2007)]{BAES07}
%Baes M., Sil'chenko O.K., Moiseev A.V. et al., 2007, A\&A, 467,
%991
\bibitem[Badenes et al.(2009)]{BADENES09}
Badenes C., Harris J., Zaritsky D. et al., 2009, ApJ, 700, 727
%\bibitem[\protect\citeauthoryear{Boissier \& Prantzos}{1999}]{BP99}
%Boissier S., Prantzos N., 1999, MNRAS, 307, 857
%\bibitem[Benetti et al.(2006)]{BENETTI06}
%Benetti S., Cappellaro E., Turatto M. et al., 2006, ApJ, 653, L129
%\bibitem[Blondin et al.(2009)]{BLONDIN09}
%Blondin S., Prieto J.L., Patat F. et al., 2009, ApJ, 693, 207
%\textbf{\bibitem[\protect\citeauthoryear{Blanc \&
%Greggio}{2008}]{BLANC08} Blanc G., Greggio L., 2008, arXiv:
%0803.3793   }
%\bibitem[Branch(1992)]{BRA92}
%Branch D., 1992, ApJ, 392, 35
%\bibitem[Branch \& Bergh(1993)]{BB93}
%Branch D., Bergh S.V., 1993, AJ, 105, 2231
%\bibitem[Branch et al.(1995)]{BRANCH95}
%Branch D., Livio M., Yungelson L.R. et al., 1995, PASP, 107, 1019
%\bibitem[Branch(2004)]{BRA04}
%Branch D., 2004, Nature, 431, 1044
%\bibitem[Branch(2006)]{BRA06}
%Branch D., 2006, Nature, 443, 283
\bibitem[Brandt et al.(2010)]{BRANDT10}
Brandt T.D. et al., 2010, AJ, 140, 804
%\bibitem[Cappellaro et al.(1997)]{CAP97}
%Cappellaro E., Turatto M., Tsvetkov D.Y., Bartunov O.S., Pollas
%C., Evans R., Hamuy M., 1997, A\&A, 322,431
%\bibitem[Cappellaro \& Turatto(1997)]{CT97}
%Cappellaro E., Turatto M., 1997, in Ruiz-Lapuente P., Cannal R.,
%Isern J., eds, Thermonuclear Supernovae. Kluwer, Dordrecht, p. 77
%\textbf{\bibitem[\protect\citeauthoryear{Chugai}{1986}]{CHU86}
%Chugai N.N., 1986, SvA, 30, 563                       }
%\bibitem[Chugai \& Yungelson(2004)]{CHU04}
%Chugai N.N., Yungelson L.R., 2004, Astronomy Letters, 30, 65
%\bibitem[Chen \& Han(20020]{CHE02}
%Chen X., Han, Z., 2002, MNRAS, 335, 948
%\bibitem[Chen \& Han(2003)]{CHE03}
%Chen X., Han, Z., 2003, MNRAS, 341, 662
\bibitem[Chen \& Tout(2007)]{CHE07}
Chen X., Tout C.A., 2007, ChJAA, 7, 2, 245
\bibitem[Chen \& Li(2007)]{CHENWC07}
Chen W., Li X., 2007, ApJ, 658, L51
\bibitem[Chen \& Li(2009)]{CHENWC09}
Chen W., Li X., 2009, ApJ, 702, 686
%\bibitem[Chen \& Han(2008)]{CHE08}
%Chen X., Han, Z., 2008, MNRAS, 387, 1416, arXiv: 0804.2294
%\bibitem[\protect\citeauthoryear{Dahlen et al.}{2004}]{DAH04}
%Dahlen T. et al., 2004, ApJ, 613, 189
%\bibitem[\protect\citeauthoryear{Delgado \& Thomas}{1981}]{DT81}
%Delgado A.J., Thomas H.C., 1981, A\&A, 96, 142
\bibitem[Della Valle et al.(2005)]{DEL05}
Della Valle M., Panagia N., Padovani P., Cappellaro E., Mannucc,
F., Turatto M., 2005, ApJ, 629, 750
%\bibitem[Deng et al.(2004)]{DEN04}
%Deng J., Kawabata K.S., Ohyama Y. et al., 2004, ApJ, 605, L37
%\bibitem[Deyoung \& Schmidt(1994)]{DEYOUNG94}
%Deyoung J. A. \& Schmidt R.E., 1994, ApJ, 431, L47
%\bibitem[Di Stefano \& Kong(2003)]{DIK03}
%Di Stefano R., Kong A.K.H., 2003, ApJ, 592, 884
%\bibitem[\protect\citeauthoryear{Della Valle et al.}{2005}]{DEL05}
%Della Valle M., Panagia N., Padovani P.,
%Cappellaro E., Mannucci F., Turatto M., 2005, ApJ, 629, 750
%\bibitem[Eggleton(1971)]{EGG71}
%Eggleton P.P., 1971, MNRAS, 151, 351
%\bibitem[Eggleton(1972)]{EGG72}
%Eggleton P.P., 1972, MNRAS, 156, 361
%\bibitem[Eggleton(1973)]{EGG73}
%Eggleton P.P., 1973, MNRAS, 163, 279
%\bibitem[Eggleton et al.(1989)]{EGG89}
%Eggleton P. P., Tout C. A., Fitechett M. J., 1989, ApJ, 347, 998
%\bibitem[Ergma, Fedorova \& Yungelson(2001)]{ERGMA01}
%Ergma E., Fedorova A.V., Yungelson L.R., 2001, A\&A, 376, L9
%\textbf{\bibitem[\protect\citeauthoryear{Filippenko}{1997}]{FIL97}
%Filippenko A.V., 1997, ARA\&A, 35, 309
\bibitem[F\"{o}rster et al.(2006)]{FORSTER06}
F\"{o}rster F., Wolf C., Podsiadlowski P., Han Z., 2006, MNRAS,
368, 1893
%\textbf{\bibitem[\protect\citeauthoryear{F\"{o}rster et
%al.}{2006}]{FORSTER06} F\"{o}rster F., Wolf C., Podsiadlowski P.,
%Han Z., 2006, MNRAS, 368, 1893}
%\bibitem[2005]{FUH05}
%Fuhrmann K., 2005, MNRAS, 359, L35
%\textbf{\bibitem[\protect\citeauthoryear{Gal-Yam \& Maoz}{2004}]{GALYAM04}
%Gal-Yam A., Maoz D., 2004, MNRAS, 347, 942
\bibitem[Graham et al.(2010)]{GRAHAM10}
Graham M.L., Pritchet C.J., Sullivan M. et al., 2010, AJ, 139, 594
\bibitem[Graur et al.(2011)]{GRAUR11}
Graur O., Poznanski D., Maoz D. et al., 2011, MNRAS, submitted,
arXiv: 1102.0005
%\bibitem[Guo et al.(2008)]{GUO08}
%Guo W., Zhang F., Meng X., Li Z., Han Z., 2008, ChJAA, 8, 63
%\bibitem[\protect\citeauthoryear{Gallagher et al.}{2005}]{GAL05}
%Gallagher J.S., Garnavich P.M., Berlind P., Challis P., Jha S.,
%Kirshner, 2005, ApJ, 634, 210
%\bibitem[Geier et al.(2007)]{GEI07}
%Geier S., Nesslinger S., Heber U., Przybilla N., Napiwotzki R.,
%Kudritzki R.-P., 2007, A\&A, 464, 299
%\bibitem[Girardi et al.(2000)]{GIRARDI00}
%Girardi L., Bressan A., Bertelli G., 2000, A\&A, 141, 371
%\bibitem[Goldberg \& Mazeh(1994)]{GM94}
%Goldberg D., Mazeh T., 1994, A\&A, 282, 801
%\textbf{\bibitem[\protect\citeauthoryear{Greggio et
%al.}{2008}]{GREGGIO08} Greggio L., Renzini A., Daddi E., 2008,
%accepted by MNRAS, arXiv: 0805.1512}
%\bibitem[\protect\citeauthoryear{Germany et al.}{2000}]{GER00}
%Germany L.M., Reiss D.J., Sadler E.M. et al., 2000, ApJ, 533, 320
%\bibitem[Greiner \& van Teeseling(1998)]{GvT98}
%Greiner J. \& van Teeseling A., 1998, A\&A, 118, 217
\bibitem[Hachisu et al.(1996)]{HAC96}
Hachisu I., Kato M., Nomoto K., ApJ, 1996, 470, L97
\bibitem[Hachisu et al.(1999a)]{HAC99a}
Hachisu I., Kato M., Nomoto K., Umeda H., 1999a, ApJ, 519, 314
%\bibitem[Hachisu et al.(1999b)]{HAC99b}
%Hachisu I., Kato M., Nomoto K., 1999b, ApJ, 522, 487
%\bibitem[Hachisu et al.(2000a)]{HAC00a}
%Hachisu I., Kato M., Kato T., Matsumoto K., 2000a, ApJ, 528, L97
%\bibitem[Hachisu et al.(2000b)]{HAC00b}
%Hachisu I., Kato M., Kato T., Matsumoto K., Nomoto K., 2000b, ApJ,
%534, L189
%\bibitem[Hachisu \& Kato(2000)]{HK00}
%Hachisu I., Kato M., 2000, ApJ, 540, 447
%\bibitem[Hachisu \& Kato(2001)]{HK01}
%Hachisu I., Kato M., 2001, ApJ, 558, 323
%\bibitem[Hachisu \& Kato(2003a)]{HK03a}
%Hachisu I., Kato M., 2003a, ApJ, 588, 1003
%\bibitem[Hachisu \& Kato(2003b)]{HK03b}
%Hachisu I., Kato M., 2003b, ApJ, 590, 445
%\bibitem[Hachisu \& Kato(2003c)]{HK03c}
%Hachisu I., Kato M., 2003c, ApJ, 598, 527
%\bibitem[Hachisu, Kato \& Schaefer(2003)]{HKS03}
%Hachisu I., Kato M., Schaefer B.E., 2003, ApJ, 584, 1008
%\bibitem[Hachisu \& Kato(2005)]{HK05}
%Hachisu I., Kato M., 2005, ApJ, 631, 1094
%\bibitem[Hachisu \& Kato(2006a)]{HK06a}
%Hachisu I., Kato M., 2006a, ApJ, 642, L52
%\bibitem[Hachisu \& Kato(2006b)]{HK06b}
%Hachisu I., Kato M., 2006b, ApJ, 651, L141
%\bibitem[Hachisu et al.(2007)]{HKL07}
%Hachisu I., Kato M., Luna G.J.M., 2007, ApJ, 659, L153
\bibitem[Hachisu et al.(2008)]{HKN08}
Hachisu I., Kato M., Nomoto K., 2008, ApJ, 679, 1390
%\bibitem[\protect\citeauthoryear{Hamuy et al.}{1995}]{HAM95}
%Hamuy M., Phillips M.M., MAZA J. ,Suntzeff N.B., Schommer R.A.,
%Avil\'{e}s R., 1995, AJ, 109, 1
%\bibitem[Hamuy et al.(1996)]{HAM96}
%Hamuy M., Phillips M.M., Schommer R.A., Schommer R.A.,Suntzeff
%N.B., Maza J., Avil\'{e}s R., 1996, AJ, 112, 2391
%\bibitem[Hamuy et al.(1996b)]{HAM96b}
%Hamuy M., Phillips M.M., Suntzeff N.B, Schommer R.A. 1996b, AJ,
%112, 2398
%\bibitem[Hamuy et al.(2003)]{HAM03}
%Hamuy M. et al., 2003, Nature, 424, 651
%\bibitem[Han, Podsiadlowski \& Eggleton(1994)]{HAN94}
%Han Z., Podsiadlowski P., Eggleton P.P., 1994, MNRAS, 270, 121
%\bibitem[Han, Podsiadlowski \& Eggleton(1995)]{HAN95}
%Han Z., Podsiadlowski P., Eggleton P.P., 1995, MNRAS, 272, 800
%\bibitem[Han(1998)]{HAN98}
%Han Z., 1998, MNRAS, 296, 1019
%\bibitem[Han et al.(2000)]{HAN00}
%Han Z., Tout C.A., Eggleton P.P., 2000, MNRAS, 319, 215
%\bibitem[Han et al.(2002)]{HAN02}
%Han Z., Podsiadlowski Ph., Maxted P. F. L., Marsh T. R., Ivanova
%N., 2002, MNRAS, 336, 449
\bibitem[Han \& Podsiadlowski(2004)]{HAN04}
Han Z., Podsiadlowski Ph., 2004, MNRAS, 350, 1301
%\bibitem[Han \& Podsiadlowski(2006)]{HAN06}
%Han Z., Podsiadlowski Ph., 2006, MNRAS, 368, 1095
%\bibitem[Hern\'{a}ndez et al.(2009)]{HERNANDEZ09}
%Hern\'{a}ndez J.I.G., Ruiz-lapuente P., Filippenko A.V., Foley
%R.J., Gal-Yam A., Simon J.D., 2009, ApJ, 691, 1
%\bibitem[Herbig et al.(1965)]{HERBIG65}
%Herbig G.H., Preston G.W., Smak J., Paczy$\acute{\rm n}$ski B.,
%1965, ApJ, 141, 617
\bibitem[Hillebrandt \& Niemeyer(2000)]{HN00}
Hillebrandt W., Niemeyer J.C., 2000, ARA\&A, 38, 191
%\bibitem[Hjellming \& Webbink(1987)]{HW87}
%Hjellming M.S., Webbink R.F., 1987, ApJ, 318, 794
%\textbf{\bibitem[\protect\citeauthoryear{Hopkins \& Beacom}{2006}]{HOPKINS06}
%Hopkins A.M., Beacom J.F., 2006, ApJ, 651, 142                       }
%\bibitem[Howell et al.(2006)]{HOW06}
%Howell D.A. et al., 2006, Nature, 443, 308
\bibitem[Hoyle \& Fowler(1960)]{HF60}
Hoyle F. \& Fowler W.A., 1960, ApJ, 132, 565
\bibitem[Hurley et al.(2000)]{HUR00}
Hurley J.R., Pols O.R., Tout C.A., 2000, MNRAS, 315, 543
\bibitem[Hurley et al.(2002)]{HUR02}
Hurley J.R., Tout C.A., Pols O.R., 2002, MNRAS, 329, 897
\bibitem[Iben \& Tutukov(1984)]{IT84}
Iben I., Tutukov A.V., 1984, ApJS, 54, 335
%\bibitem[Iglesias \& Rogers(1996)]{IR96}
%Iglesias C. A., Rogers F. J., 1996, ApJ, 464, 943
%\bibitem[Ihara et al.(2007)]{IHA07}
%Ihara Y., Ozaki J., Doi M. et al., 2007, PASJ, 59, 811, arXiv:
%0706.3259
%\bibitem[Kato \& Hachisu(2004)]{KH2004}
%Kato M., Hachisu I., 2004, ApJ, 613, L129
\bibitem[Khan et al.(2010)]{KHAN10}
Khan R., Prieto J.L.  Pojma\'{n}ski G. et al., 2011, ApJ, 726, 106
%\bibitem[Kippenhahn \& Weigert(1967)]{KW67}
%Kippenhahn R., Weigert A., 1967, ZA, 65, 251
\bibitem[Kobayashi et al.(1998)]{KOB98}
Kobayashi C., Tsujimoto T., Nomoto K. et al., 1998, ApJ, 503, L155
%\bibitem[Kotak et al.(2004)]{KOT04}
%Kotak R., Meikle W.P.S., Adamson S. et al., 2004, MNRAS, 354, L13
\bibitem[Langer et al.(2000)]{LAN00}
Langer N., Deutschmann A., Wellstein S. et al., 2000, A\&A, 362,
1046
\bibitem[Leibundgut(2000)]{LEI00}
Leibundgut B., 2000, A\&ARv, 10, 179
%\bibitem[Lemasle et al.(2007)]{LEMASLE07}
%Lemasle B., Piersimoni A., Pedicelli P. et al., 2007, arXiv:
%0711.3988
%\bibitem[Leonard(2007)]{LEO07}
%Leonard D.C., 2007, ApJ, 670, 1275
\bibitem[Li \& van den Heuvel(1997)]{LI97}
Li X.D., van den Heuvel E.P.J., 1997, A\&A, 322, L9
%\bibitem[Liebert et al.(2003)]{LIE03}
%Liebert J., Bergeron P., Holberg J.B., 2003, AJ, 125, 348
%\bibitem[Liebert et al.(2005)]{LIE05}
%Liebert J., Bergeron P., Holberg J.B., 2005, ApJS, 156, 47
%\bibitem[Livio \& Soker(1988)]{LS88}
%Livio M., Soker N., 1988, ApJ, 329, 764
%\bibitem[Livio \& Riess(2003)]{LR03}
%Livio M., Riess A., 2003, ApJ, 594, L93
\bibitem[L\"{u} et al.(2009)]{LGL09}
L\"{u} G., Zhu C. Wang Z., Wang N., 2009, MNRAS, 396, 1086
%\bibitem[Lockley et al.(1997)]{LOCKLEY97}
%Lockley J.J., Eyres S.P.S., Wood J.H., 1997, MNRAS, 287, L14
%\bibitem[Lockley et al.(1999)]{LOCKLEY99}
%Lockley J.J., Wood J.H., Eyres S.P.S., Naylor T., Shugarov S.,
%1999, MNRAS, 310, 963
\bibitem[Maoz et al.(2010)]{MAOZ10a}
Maoz D., Keren S., Avishay G., 2010, ApJ, 722, 1879
\bibitem[Maoz (2010)]{MAOZ10b}
Maoz D., 2010, AIPC, 1314, 223
\bibitem[Maoz et al.(2011)]{MAOZ10}
Maoz D., Mannucci F., Li, W. et al., 2011, MNRAS, 412, 1508
\bibitem[Mannucci et al.(2005)]{MAN05}
Mannucci F., Della Valle M., Panagia N., Cappellaro E., Cresci G.,
Maiolino R., Petrosian A., Turatto M., 2005, A\&A, 433, 807
\bibitem[Mannucci et al.(2006)]{MAN06}
Mannucci F., Della Valle M., Panagia N., 2006, MNRAS, 370, 773
%\bibitem[Marietta et al.(2000)]{MAR00}
%Marietta E., Burrows A., Fryxell B., 2000, ApJS, 128, 615
%\bibitem[Mattila et al.(2005)]{MAT05}
%Mattila S., Lundqvist P., Sollerman J. et al., 2005, A\&A, 443,
%649
%\bibitem[Mazeh et al.(1992)]{MAZ92}
%Mazeh T., Goldberg D., Duquennoy A., Mayor M., 1992, ApJ, 401, 265
\bibitem[Meng et al.(2006)]{MEN06}
Meng X., Chen X., Tout C.A., Han Z., 2006, ChJAA, 6, 4, 461
%\bibitem[Meng et al.(2007)]{MEN07}
%Meng X., Chen X., Han Z., 2007, PASJ, 59, 835
%\bibitem[Meng, Chen \& Han(2008)]{MEN08}
%Meng X., Chen X., Han Z., 2008, A\&A, 487, 625
\bibitem[Meng, Chen \& Han(2009)]{MENG09}
Meng X., Chen X., Han Z., 2009, MNRAS, 395, 2103
\bibitem[Meng \& Yang(2010a)]{MENGYANG10a}
Meng X., Yang W., 2010a, ApJ, 710, 1310
\bibitem[Meng \& Yang(2010b)]{MENGYANG10b}
Meng X., Yang W., 2010b, A\&A, 516, A47
%\bibitem[Mennickent \& Honeycutt(1995)]{MH95}
%Mennickent R.E. \& Honeycut R.K., 1995, Inf. Bull.Variable Stars,
%4232
\bibitem[Mennekens et al.(2010)]{Mennekens10}
Mennekens N., Vanbeveren D., De Greve J. P., De Donder, E., 2010,
A\&A, 515, A89
%\textbf{\bibitem[\protect\citeauthoryear{Meng et
%al.}{2008a}]{MEN08a} Meng X., Chen X., Han Z., 2008a, A\&A, in
%preparation} \textbf{\bibitem[\protect\citeauthoryear{Meng et
%al.}{2008a}]{MEN08b} Meng X., Chen X., Han Z., 2008b, MNRAS, in
%preparation}
\bibitem[Miller \& Scalo(1979)]{MS79}
Miller G.E., Scalo J.M., 1979, ApJS, 41, 513
\bibitem[Nagamine et al.(2001)]{NAAGAMINE01}
Nagamine K., Fukugita M., Cen R. et al., 2001, ApJ, 558, 497
%\bibitem[Napiwotzki et al.(2004)]{NAPIWOTZKI04}
%Napiwotzki R., Karl C., Nelemans G. et al., 2004, RMxAC, 20, 113
%\bibitem[Nelemans et al.(2000)]{NELEMANS00}
%Nelemans G., Verbunt F., Yungelson L.R. et al., 2000, A\&A, 360,
%1011
%\bibitem[Nelemans \& Tout(2005)]{NELEMANS05}
%Nelemans G., Tout C.A., 2005, MNRAS, 356, 753
\bibitem[Nomoto, Thielemann \& Yokoi(1984)]{NTY84}
Nomoto K., Thielemann F-K., Yokoi K., 1984, ApJ, 286, 644
%\bibitem[Nomoto \& Kondo (1991)]{NK91}
%Nomoto K., Kondo Y., 1991, ApJ, 367, L19
\bibitem[Nomoto et al.(1999)]{NOM99}
Nomoto K., Umeda H., Hachisu I. Kato M., Kobayashi C., Tsujimoto
T., 1999, in Truran J., Niemeyer T., eds, Type Ia Suppernova
:Theory and Cosmology.Cambridge Univ. Press, New York, p.63
%\bibitem[Nomoto et al.(2003)]{NOM03}
%Nomoto K., Uenishi T., Kobayashi C. Umeda H., Ohkubo T., Hachisu
%I., Kato M., 2003, in Hillebrandt W., Leibundgut B., eds, From
%Twilight to Highlight: The Physics of supernova, ESO/Springer
%serious ``ESO Astrophysics Symposia'' Berlin: Springer, p.115
%\bibitem[Ofek et al.(2007)]{OFE07}
%Ofek E.O., Cameron P.B., Kaslwal M.M. et al., 2007, ApJ, 659, L13,
%arXiv: 0612408
\bibitem[Oda et al.(2008)]{ODA08}
Oda T., Totani T., Yasuda N. et al., 2008, PASJ, 60, 169
%\bibitem[Paczy$\acute{\rm n}$ski(1976)]{PAC76}
%Paczy$\acute{\rm n}$ski B., 1976, in Eggleton P.P., Mitton S.,
%Whelan J., eds, Structure and Evolution of Close Binaries. Kluwer,
%Dordrecht, p. 75
%\bibitem[Pakull et al.(1993)]{PAKULL93}
%Pakull M.W., Moch C., Bianchi L., Thomas H.C., Guibert J.,
%Beaulieu J.P., Grison P., \& Schaeidt S., 1993, A\&A, 278, L39
%\bibitem[Paresce et al.(1995)]{PARESCE95}
%Paresce F., Livio M., Hack W., Korista K., 1995, A\&A, 299, 823
\bibitem[Parthasarathy et al.(2007)]{PAR07}
Parthasarathy M., Branch D., Jeffery D.J., Baron E., 2007, NewAR,
51, 524, arXiv: 0703415
%\bibitem[Patat et al.(2007)]{PAT07}
%Patat E. et al., Science, 317, 924
%\bibitem[Patterson et al.(1998)]{PATTERSON98}
%Patterson J., Kemp J., Shambrook A. et al., 1998, PASP, 110, 380
%\bibitem[Podsiadlowski et al.(2002)]{POD02}
%Podsiadlowski P., Rappaport S., Pfahl, 2002, ApJ, 565, 1107
%\bibitem[Podsiadlowski et al.(2006)]{POD06}
%Podsiadlowski P., Mazzali P.A., Lesaffre P., Wolf C., F\"{o}rster
%F., 2006, arXiv: 0608324
\bibitem[Perlmutter et al.(1999)]{PER99}
Perlmutter S. et al., 1999, ApJ, 517, 565
%\bibitem[Phillips(1993)]{PHI93}
%Phillips M.M., 1993, ApJ, 413, L105
%\bibitem[Pols et al.(1995)]{POL95}
%Pols O.R., Tout C.A., Eggleton P.P. et al., 1995, MNRAS, 274, 964
%\bibitem[Pols et al.(1997)]{POL97}
%Pols O.R., Tout C.A., Schr\"{o}der K.P. et al., 1997, MNRAS, 289,
%869
%\bibitem[Pols et al.(1998)]{POL98}
%Pols O.R., Schr\"{o}der K.P., Hurly J.R. et al., 1998, MNRAS, 298,
525
\bibitem[Prieto et al.(2008)]{PRIETO08}
Prieto J.L. et al., 2008, ApJ, 673, 999
%\bibitem[Prieto et al.(2007)]{PRI07b}
%Prieto J.L. et al., 2007, arXiv: 0706.4088
%\bibitem[Quimby, H\"{o}flich \& Wheeler(2007)]{QUI07}
%Quimby R., P. H\"{o}flich, J.C. Wheeler, 2007, ApJ, 666, 1083
\bibitem[Raskin(2009)]{RASKIN09}
Raskin C., Scannapieco E., Rhoads J., Della Valle, M., 2009, ApJ,
707, 74
%\bibitem[Retter, Leibowitz \& Ofek(1997)]{RETTER97}
%Retter A., Leibowitz E.M. \& Ofek E.O., 1997, MNRAS, 286, 745
%\bibitem[\protect\citeauthoryear{Rigon et al.}{2003}]{RIG03}
%Rigon L. et al., 2003, MNRAS, 340, 191
%\bibitem[Riess et al.(1999)]{RIE99}
%Riess A. et al., 1999, AJ, 117, 707
\bibitem[Riess et al.(1998)]{RIE98}
Riess A. et al., 1998, AJ, 116, 1009
%\bibitem[\protect\citeauthoryear{Roberts \& Haynes}{1994}]{RH94}
%Roberts M.S., Haynes M.P., 1994, ARA\&A, 32, 115
%\textbf{\bibitem[\protect\citeauthoryear{Roelofs et
%al.}{2008}]{ROELOFS08} Roelofs G., et al. 2008, MNRAS, submitted
%(arXiv: 0802.2097)}
%\bibitem[\protect\citeauthoryear{Ruiz-Lapuente et al.}{1995}]{RUI95}
%Ruiz-Lapuente P., Burkert A., Canal R., 1995, ApJ, 447, L69
%\bibitem[Ruiz-Lapuente et al.(2004)]{RUI04}
%Ruiz-Lapuente P. et al., 2004, Nature, 431, 1069
\bibitem[Scannapieco \& Bildsten(2005)]{SB05}
Scannapieco E., Bildsten L., 2005, ApJ, 629, L85
%\bibitem[Schr\"{o}der et al.(1997)]{SCH97}
%Schr\"{o}der K.P., Pols O.R., Eggleton P.P., 1997, MNRAS, 285, 696
\bibitem[Schawinski(2009)]{SCHAWINSKI09}
Schawinski K., 2009, MNRAS, 397, 717
%\bibitem[Schaefer \& Ringwald(1995)]{SR95}
%Schaefer B.E. \& Ringwald F.A., 1995, ApJ, 447, L45
%\bibitem[Schaefer(1990)]{SCHAEFER90}
%Schaefer B.E., 1990, ApJ, 355, L39
%\bibitem[Shanks et al.(2002)]{SHA02}
%Shanks T., Allen P.D., Hoyle F. et al., 2002, ASPC, 283, 274
\bibitem[Strolger et al.(2004)]{STROLGER04}
Strolger L.-G. et al., 2004, ApJ, 613, 200
\bibitem[Strolger et al.(2010)]{STROLGER10}
Strolger L.-G. et al., 2010, ApJ, 713, 32
%\bibitem[Steiner \& Diaz (1998)]{STEINER98}
%Steiner J.E. \& Diaz M.P., 1998, PASP, 110, 276
\bibitem[Sullivan et al.(2006)]{SULLIVAN06}
Sullivan M. et al. 2006, ApJ, 648, 868
%\bibitem[Timmes et al.(1997)]{TIM97}
%Timmes F.X., Diehl R., Hartmann D.H., 1997, ApJ, 479, 760
%\bibitem[Timmes et al.(20030]{TIM03}
%Timmes F.X., Brown E.F., Truran J.W., 2003, ApJ, 590, L83
%\bibitem[\protect\citeauthoryear{Tout et al.}{1997}]{TOU97}
%Tout C.A., Aarseth S.J., Pols O.R., Eggleton P.P., 1997, MNRAS,
%291, 732
%\bibitem[Travaglio et al.(2005)]{TRA05}
%Travaglio C., Hillebrandt W., Reinecke M., 2005, A\&A, 443, 1007
%\bibitem[\protect\citeauthoryear{Turatto et al.}{2000}]{TUR00}
%Turatto M. et al., ApJ, 534, L57
%\bibitem[\protect\citeauthoryear{Terlevich \& Forbes}{2002}]{TF02}
%Terlevich A.I., Forbes D.A., 2002, MNRAS, 330, 547
%\bibitem[Tutukov \& Yungelson(2002)]{TUT02}
%Tutukov A.V., Yungelson L.R., 2002, Astron. Rep., 46, 667
\bibitem[Totani et al.(2008)]{TOTANI08}
Totani T., Morokuma T., Oda T. et al., 2008, PASJ, 60, 1327
%arXiv: 0804.0909
%\bibitem[Uenishi et al.(2004)]{UEN04}
%Uenishi T., Suzuki T., Nomoto K. et al., 2004, Rev. Mex. A\&A, 20,
%219
\bibitem[Umeda et al.(1999a)]{UME99}
Umeda H., Nomoto K., Yamaoka H. et al., 1999a, ApJ, 513, 861
\bibitem[Umeda et al.(1999b)]{UME99b}
Umeda H., Nomoto K., Kobayashi C. et al., 1999b, ApJ, 522, L43
\bibitem[Valiante et al. (2009)]{VALIANTE09}
Valiante R., Matteucci F., Recchi S., Calura F., 2009, NewA, 14,
638
%\bibitem[van den Bergh \& Tammann(1991)]{VAN91}
%van den Bergh S., Tammann G.A., 1991, ARA\&A, 29, 363
%\bibitem[Voss \& Nelemans(2008)]{VOSS08}
%Voss R., \& Nelemans G., 2008, Nature, 451, 802
%\bibitem[Wang et al.(1997)]{WAN97}
%%Wang L., H\"{o}flich P., Wheeler J.C., 1997, ApJ, 483, L29
%\bibitem[Wang et al.(2004)]{WAN04}
%Wang L., Baade D., H\"{o}flich P. et al., 2004, ApJ, 604, L53
\bibitem[Wang et al.(2009)]{WANGB09a}
Wang B., Meng X., Chen X., Han, Z., 2009, MNRAS, 395, 847
%\bibitem[\protect\citeauthoryear{Wang et al.}{2009b}]{WANGB09b}
%Wang, B., Chen, X., Meng, X., Han, Z., 2009b, ApJ, 701, 1540
\bibitem[Wang, Li \& Han(2010)]{WANGB10}
Wang B., Li X., Han Z., 2010, MNRAS, 401, 2729
%\bibitem[Webbink(1988)]{WEBBINK88}
%Webbink R. F., 1988, in The Symbiotic Phenomenon, eds. J.
%Mikolajewska, M. Friedjung, S. J. Kenyon \& R. Viotti (Kluwer:
%Dordrecht), p.311
%\bibitem[Webbink(2007)]{WEBBINK07} Webbink R., 2007, arXiv:
%0704.0280
\bibitem[Webbink(1984)]{WEB84}
Webbink, R.F., 1984, ApJ, 277, 355
\bibitem[Whelan \& Iben(1973)]{WI73}
Whelan J., Iben I., 1973, ApJ, 186, 1007
%\bibitem[Whelan \& Iben(1987)]{WI87}
%Whelan J., Iben I., 1987, in Philipp A.G.D., Hayes D.S., Liebert
%J.W., eds, IAU Colloq.95, Second Conference on Faint Blue Stars.
%Davis Press, Schenectady, p. 445
%\bibitem[Wickramasinghe \& Ferrario(2005)]{WF05}
%Wickramasinghe D.T., Ferrario L., 2005, MNRAS, 356, 1576
%\bibitem[Willems \& Kolb(2004)]{WK04}
%Willems B., Kolb U., 2004, A\&A, 419, 1057
%\bibitem[Wood-Vasey et al.(2004)]{WOO04}
%Wood-Vasey W.M., Wang L., Aldering G., 2004, ApJ, 616, 339
%\bibitem[Wood-Vasey \& Sokoloski(2006)]{WOO06}
%Wood-Vasey W.M., Sokoloski J.L., 2006, ApJ, 645, L53
\bibitem[Xu \& Li(2009)]{XL09}
Xu X. \& Li X., 2009, A\&A, 495, 243
%\bibitem[Yungelson et al.(1995)]{YUN95}
%Yungelson L., Livio M., Tutukou A. Kenyon S.J., 1995, ApJ, 447,
%656
%\bibitem[Yungelson \& Livio(1998)]{YUN98}
%Yungelson L., Livio M., 1998, ApJ, 497, 168
%\bibitem[Yungelson \& Livio(2000)]{YUN00}
%Yungelson L., Livio M., 2000, ApJ, 528, 108

\end{thebibliography}
\end{document}